\documentstyle[array,psfig,aps,float]{revtex}
\newcommand{\bm}[1]{\mbox{\protect\boldmath $#1$}}
\begin{document}
\draft
\title{Propagation of mesons in asymmetric nuclear matter in a density dependent coupling model}
\author{R. Aguirre and A.L. De Paoli}
\address{Department of Physics, La Plata National University\\
C.C. 67 (1900) La Plata, Argentina }
\maketitle
\begin{abstract}
We study the propagation of the light mesons $\sigma, \omega, \rho$,
and $a_0$(980) in dense hadronic matter in an extended derivative
scalar coupling model. Within the scheme proposed it is possible to
unambiguously define effective density-dependent couplings at the
lagrangian level. We first apply the model to study asymmetric nuclear
matter with fixed isospin asymmetry, and then we pay particular
attention to hypermatter in $\beta$-equilibrium. The equation of state
and the potential contribution to the symmetry coefficient arising from
the mean field approximation are investigated.
\end{abstract} 

\section{Introduction}
\label{0} In recent years the medium dependence of the meson-baryon
couplings has been object of speculation
\cite{TOKI,WEIGEL,LENSKE1,LENSKE2,LENSKE3,TYPEL,BANERJEE,RAKHIMOV,THOMAS}.
This subject has been promoted by the successes of the so-called
quantum hadrodynamics theory (QHD) \cite{WALECKA}, in the relativistic
description of diverse nuclear phenomena.

The assumption of variable couplings in the mean field approximation
(MFA) is founded on different grounds. It can be interpreted as the
trace of the quark structure of hadrons
\cite{BANERJEE,RAKHIMOV,THOMAS}, or it can be viewed as a way to match
effective lagrangians and free-space nucleon-nucleon interactions
\cite{TOKI,WEIGEL,LENSKE1,LENSKE2,LENSKE3,TYPEL}. In any case assigning
a variable behavior to the couplings seems to be an appropriate method
to interpolate from one dynamical regime to another, using effective
hadron field models. A similar meaning has been given to the in-medium
meson masses, which have been related to the transition to the chiral
regime \cite{BROWN-RHO}. The so-called Brown-Rho scaling law
qualitatively describes the behaviour of the hadronic masses in the
proximity of the transition point. According to the scaling law, all
hadronic masses decrease approximately at the same rate as the system
approaches to the chiral phase transition (with exception of the
pseudo-scalar meson masses). Applied to the light vector mesons this
hypothesis could explain some experimental results, as for example the
dilepton production rate in heavy ion collisions.

On the other hand, in certain purely hadronic models
\cite{FELDMEIER,Z&M} the non-polynomial meson-nucleon interaction gives
rise to effective, density-dependent coupling in the MFA
\cite{BHATTAC,AGUIRRE,AGUIRRE2}. In this paper we propose an extension
of the derivative scalar coupling model (DSCM) of \cite{Z&M}, which
preserves the charge symmetry and provides effective couplings for all
the mesonic channels. The medium dependence comes through the mean
value of the scalar $\sigma$ meson, evaluated at finite density and
temperature. More general interactions could include non-linear
vertices in terms of all the mesons considered, but we restrict here to
the simplest case. This choice is similar to the approach of
\cite{LENSKE1}, where the functional dependence of the couplings
includes only the product $\bar{\Psi}\gamma_\mu \Psi$, but not the
scalar $\bar{\Psi} \Psi$ or more involved nucleon field combinations.
 As stressed in this reference, the density variation of the
couplings must be written as a Lorentz invariant functional of the
considered fields, to obtain the correct Euler-Lagrange equations.
Otherwise the so called ``rearrangement" contribution is absent
and the thermodynamical consistency is lost.

We use the model proposed here to investigate asymmetric nuclear matter
at finite temperature, therefore we explicitly include the isovector
mesons $\rho$ and $a_0(980)$.

The equation of state of asymmetric matter is an important input in
astrophysical studies such as the cooling rate of neutron stars or the
supernova collapse mechanism. Also it is of interest in the description
of heavy-ion collisions, where experiences with radioactive beams are
expected to provide new insights into the structure of dense matter
with high degree of asymmetry. This situation has stimulated the
interest in this issue, and many theoretical works have been done in
the last years
\cite{LENSKE2,LI1,LI2,KUTSCHERA,KUTSCHERA1,KUTSCHERA2,KUTSCHERA3,BROWN,VARIOS,GRECO}.

As an important case of asymmetric matter we consider hadronic matter
in $\beta$-equilibrium. For this purpose we generalize the DSCM  model
to include the octet of baryons $n, \Lambda, \Sigma$, and $\Xi$. It is
well known that the proton concentration is determinant in the cooling
of neutron stars \cite{LATTIMER,LATTIMER2}, and this concentration is
mainly determined by the isospin-dependent contribution to the equation
of state.

It is also of interest to study the meson properties under these
conditions since, for instance, it could carry information about
transient states of matter during heavy-ion collisions. This subject
has been poorly developed in the literature. We have evaluated the
meson propagators in the relativistic random phase approximation
(RRPA), including particle-antiparticle contributions, and we have
extracted from them the effective meson masses.

We organize this paper presenting the model in Sec. \ref{SEC2}, in Sec.
\ref{SEC3} we discuss the bulk properties of symmetric nuclear matter
at zero temperature, meanwhile the asymmetric nuclear matter equation
of state is treated in Sec. \ref{SEC4}. The properties of
$\beta$-stable matter are considered in Sec. \ref{SEC5}, and the
Feynman graphs contributing to the RRPA, the evaluation of the
propagators and the behavior of mesons in the hadronic environment are
presented in Sec. \ref{SEC6}. We conclude with the discussion and
summary in Sec. \ref{SEC7}.

\section{The modified DSCM}
\label{SEC2}

In this section we present a relativistic model of hadronic fields
inspired on the DSCM proposed by Zimanyi and Moszkowski \cite{Z&M}.
 The DSCM has been used to study nuclear many-body
effects in several applications \cite{VARIOS2}, to investigate neutron
star properties \cite{GLENDENNING}, extended to include nucleon
resonances \cite{CHOUD} and hyperons \cite{BARRANCO},
 related to an effective quark description of hadronic properties
\cite{SCHVE}, and generalized with a tensor coupling \cite{BIRO}
in order to improve the spin-orbit splitting.

The DSCM has two important features which distinguish it from the QHD-I
model of Ref. \cite{WALECKA}. In first place it is non-renormalizable
{\it ab initio} and there is no immediate way to introduce vacuum
corrections to the MFA to the ground state, although the main
properties of nuclear matter are successfully described. In second
place a residual interaction can be extracted beyond the lowest order
solution, whose strength decreases monotonically as a function of the
baryonic density \cite{AGUIRRE,AGUIRRE2}. This fact ensures the ground
state predominance at high density as assumed in QHD \cite{WALECKA}.

Since we are interested here in the description of asymmetric matter,
besides the fields $\Psi^a$ for the nucleons, we include the isoscalar
scalar ($\sigma$) and isoscalar vector ($\omega_\mu$) mesonic fields,
and those corresponding to the $\rho$ isovector vector ($\rho_\mu^A$)
and the $a_0 (980)$ isovector scalar ($\delta^A$) mesons. We use greek,
latin lowercase and latin uppercase indexes to denote Lorentz, baryon
isospin and meson isospin components, respectively.

In its simplest version the DSCM \cite{Z&M} has a Yukawa type
N-$\omega$ coupling and a N-$\sigma$ non-polynomic term. We modify the
vertices allowing for two different mesons (one of them the scalar
$\sigma$) to locally interact with a baryon:

\begin{eqnarray}
{\cal L}_{DSC}&=& \bar{\Psi}\left[i \not \! \partial -\frac{M - g_d
\,{\bm \tau} \cdot {\bm\delta}+ g_w \not \! \omega + g_r {\bm \tau}
\cdot \not \! \! {\bm \rho}/2} {1+g_s \sigma/M} \right] \Psi +
\frac{1}{2} (\partial^\mu \sigma \partial_\mu \sigma - m_s^2 \sigma^2)+
\frac{1}{2}(\partial^\mu {\bm \delta}
\partial_\mu {\bm \delta} - m_d^2 {\bm \delta}^2)  \nonumber \\
-&\frac{1}{4}&
 F^{\mu \nu} F_{\mu \nu} + \frac{1}{2} m_w^2
\omega^2 - \frac{1}{4} {\bm R}^{\mu \nu} {\bm R}_{\mu \nu} +
\frac{1}{2} m_r^2 {\bm \rho}^2 , \label{DSCM}
\end{eqnarray}

\noindent where $\Psi (x)$ is the isospin multiplet nucleon field,
$M$ is the averaged nucleon mass and $g_s, g_d, g_v,$ and $g_r$
are adimensional coupling constants. As usual in QHD the ground
state for homogeneous infinite matter is approximated by
considering mesonic fields as classical quantities and
assimilating them to effective nucleon properties. Thus we can
separate the c-number contributions:
\begin{eqnarray}
\sigma(x) &=& \bar{\sigma}  + s(x), \label{REPLACEMENTS} \\
\delta^A (x)&=& \bar{\delta}\,\, \delta^{3A}+ d^A(x), \label{REPLACEMENTD} \\
\omega_{\mu}(x) &=&\bar{\omega} \,\, \delta_{\mu 0} + w_{\mu}(x),\\
\rho_{\mu}^A(x) &=&\bar{\rho} \,\, \delta_{\mu 0} \delta^{3A}+
r_{\mu}^A(x),\label{REPLACEMENTR}
\end{eqnarray}

\noindent where $\bar{\sigma}$, $\bar{\delta}$, $\bar{\omega}$, and
$\bar{\rho}$ are classical mean field values and $s, d^A, w_{\mu}$, and
$r^A_\mu$ are quantum fluctuations which are not included in the ground
state. Expressions for the c-number contribution to meson fields can be
obtained by taking statistical averaged Euler-Lagrange equations, and
requiring self-consistency. In this way we obtain:
\begin{eqnarray}
m_s^2 \bar{\sigma}&=&g_s  \frac{<\bar{\Psi} (M - g_d \tau_3
\bar{\delta} + g_w \gamma_0 \bar{\omega} + \frac{1}{2} g_r \tau_3
\gamma_0 \bar{\rho}) \Psi>}{M N^2} , \label{SELFCONS} \\ m_d^2
\bar{\delta}&=&g_d  \frac{<\bar{\Psi} \tau_3 \Psi>}{N},\\
 m_w^2 \bar{\omega} &=&g_w  \frac{<\Psi^{\dagger} \Psi>}{N} , \label{SELFCONSW}\\
 m_r^2 \bar{\rho} &=&g_r \frac{<\Psi^{\dagger} \tau_3 \Psi>}{N} ,\label{SELFCONSR}
\end{eqnarray}

\noindent where we have used $N=1+ g_s \bar{\sigma} / M$. The
expectation values must be evaluated with the ground state solution for
the nucleon field, which depends on $\bar{\sigma}$ and $\bar{\delta}$
through the effective nucleon mass:
\begin{equation}
M_i^{\ast}=\frac{M-g_d I_i \bar{\delta}}{N}, \label{EFFMASS}
\end{equation}
\noindent with $I_i=1,-1$ for protons and neutrons, respectively. The
nucleon dispersion relation is also modified according to $(p_0-g_w
\bar{\omega}-g_r I_i \bar{\rho}/2)^2-p^2=M^{\ast\, 2}_i$.

 In Eqs. (\ref{SELFCONSW}) and (\ref{SELFCONSR}) the terms between
 angular brackets represent the conserved baryon density and the isospin
 density, respectively.

A residual nucleon-meson interaction arises beyond the lowest
order approximation \cite{AGUIRRE} by inserting Eqs.
(\ref{REPLACEMENTS})-(\ref{REPLACEMENTR}) in the interaction term:
\begin{eqnarray}
\frac{M+ \sum_i \Gamma_i \phi_i}{1+g_s \sigma/M}&=& \frac{M+\sum_i
\Gamma_i (\bar{\phi}_i+ \delta \phi_i)}{N (1+\frac{g_s}{N} s)}.
\nonumber
\end{eqnarray}

 In the expression above the symbol $\phi_i$ represents any one of the
mesonic fields $\delta, \omega$, and $\rho$, which by virtue of
Eqs.(\ref{REPLACEMENTD})-(\ref{REPLACEMENTR}) splits into the classical
mean value $\bar{\phi_i}$, and the quantum fluctuation $\delta \phi_i$.
$\Gamma_i$ stands for the bare meson nucleon vertices:   $\Gamma_i=-g_d
{\mathbf \tau}, g_w \gamma, g_r {\mathbf \tau} \gamma/2$, corresponding
to the  $a_0, \omega$, and $\rho$ mesons respectively. The right hand
side of this equation is non-polynomic and can not be used to directly
apply a diagrammatic expansion. Restricting to the physical regime for
which quantum fluctuations are negligible compared to mean values,
enables us to approximate:

\begin{eqnarray}
\frac{M+ \sum_i \Gamma_i \phi_i}{1+g_s \sigma/M}&\simeq&M^\ast +
\gamma_0 \,\, \delta \varepsilon + {\cal L}_{res}, \label{Lres1}
\end{eqnarray}
\noindent with
\begin{eqnarray}
\delta \varepsilon&=&\sum_{\omega,\rho} \Gamma_i^\ast
\bar{\phi}_i,\label{Lres2} \\ {\cal L}_{res}&=&-g_s^\ast s +
\sum_i \Gamma_i^\ast \left[\delta \phi_i - \frac{g_s}{N M}
(\bar{\phi}_i + \delta \phi_i)s \right].\label{Lres3}
\end{eqnarray}

 We have introduced the medium dependent vertices $\Gamma_i^*$,
which are obtained from the bare ones by replacing the coupling
constants $g_d, g_w$, and $g_r$ by effective couplings. The last ones
are given by the relation $g_d/g_d^*=g_w/g_w^*=g_r/g_r^*=N$. Also we
have used $g_s/g_s^\ast=N^2$.

The expansion proposed in Eq. (\ref{Lres1}) respects the organizational
principle of nuclear effective field theories \cite{SEROT}.

In this approximation, the residual interaction of Eq.(\ref{Lres3})
arises besides the nucleon effective mass (\ref{EFFMASS}) and the
contribution to the nucleon single particle energy (\ref{Lres2}). The
interaction term ${\cal L}_{res}$ comprises a one meson-nucleon vertex,
together with a two-meson exchange term. In all cases the vertex
functions are medium dependent, $g_s^\ast\left(1+ \sum_i \Gamma_i
\bar{\phi_i} / M \right), \Gamma_d^\ast, \Gamma_w^\ast$, and
$\Gamma_r^\ast$ for the one-meson case and $g_s^\ast \Gamma_i/M$ for
the two-meson instance.

This linearized version can be used to study the quantum
corrections beyond the mean field approximation.

Variable couplings are an expected feature of hadronic models, whenever
the quark substructure becomes relevant
\cite{BANERJEE,RAKHIMOV,THOMAS}. Furthermore, density dependent
couplings have been proposed as a key assumption in order to match
relativistic nucleon potentials adjusted to scattering data, with
hadronic field models \cite{TOKI,WEIGEL,LENSKE1,LENSKE2,LENSKE3,TYPEL}.
This approach was initiated as a way to avoid involved
Brueckner-Hartree-Fock calculations for finite systems, using one boson
exchange potentials. Thus, the main purpose is to take advantage of the
relative simplicity of the Hartree approach to the QHD models. The link
between both schemes is established by requiring the equality of the
nucleon self-energy in symmetric nuclear matter as evaluated in both
formulations, and allowing QHD coupling constants to be density
dependent.

 In our treatment the effective couplings are unambiguously
extracted from the lagrangian, once the MFA has been introduced. Thus
in this scheme one has a well defined and invariant way to describe the
medium influence on the couplings. Furthermore, the internal
consistency of the approach is guaranteed.

Up to this point we have restricted the discussion only to protons and
neutrons, however the introduction of hyperons is straightforward. A
sum over different baryonic species must be considered in the
lagrangian density and the vertices must be modified in order to take
into account the isospin degeneracy of each one. Also, additional
couplings between the mesons and every hyperon iso-multiplet must be
introduced. A more detailed discussion will be given in Sec.
\ref{SEC5}.

The energy density $E$ for infinite homogeneous hadronic matter, can be
evaluated  in the MFA by taking the statistical average of the energy
momentum tensor: $E=<T^{00}>$. The thermodynamical pressure $P$ under
the same conditions is obtained by averaging the trace of the
spatial-spatial component of this tensor: $P=<$Tr$\, T^{ij}>/3$. We
include the corresponding equations for the sake of completeness:

\begin{eqnarray}
E&=&\sum_{i=p,n} \frac{1}{(2\pi)^3}\int_0^\infty d^3k E_{k\,i}
\left[ n_F(E_{k\,i})+n_F(-E_{k\,i})\right]+ \frac{1}{2}(m_s^2
{\bar{\sigma}}^2 + m_d^2 {\bar{\delta}}^2- m_w^2 {\bar{\omega}}^2
- m_r^2 {\bar{\rho}}^2) \nonumber \\ &+& g_w^\ast {\bar{\omega}}\,
n +
\frac{1}{2} g_r^\ast {\bar{\rho}}\,(n_p-n_n), \label{ENERGY} \\
P&=&\sum_{i=p,n} \frac{1}{3(2\pi)^3}\int_0^\infty d^3k
\frac{k^2}{E_{k\,i}} \left[ n_F(E_{k\,i})+n_F(-E_{k\,i})\right]-
\frac{1}{2}(m_s^2 {\bar{\sigma}}^2 + m_d^2 {\bar{\delta}}^2- m_w^2
{\bar{\omega}}^2 - m_r^2 {\bar{\rho}}^2),\label{PRESSURE}
\end{eqnarray}

\noindent with $E_{k\,i}=\sqrt{M_i^{\ast 2}+k^2}$, $M_i^\ast$ is
given by Eq. (\ref{EFFMASS}), and

\begin{eqnarray}
n_F(z_i)=\frac{\Theta(z_i)}{1+e^{\beta(x_i-\mu_i)}}+
\frac{\Theta(-z_i)}{1+e^{\beta(x_i+\mu_i)}},\nonumber
\end{eqnarray}
\noindent is the nucleon statistical occupation number, $x_i=z_i+g_w
\bar{\omega}+I_i g_r \bar{\rho}/2$ is the particle energy, $\mu_i$ is
the chemical potential, and $\beta=1/k_B T$. The chemical potential is
related to the number density of the i-type particle through:
\begin{equation}
n_i=\frac{1}{(2 \pi)^3}\int_0^{\infty} d^3k \left[
n_F(E_{k\,i})-n_F(-E_{k\,i})\right],\label{IPARTICLE}
\end{equation}
\noindent and finally $n=n_p+n_n$ is the total particle number
density.

To evaluate Eqs. (\ref{ENERGY}) and (\ref{PRESSURE}), one must fix
the particle number densities $n_p, n_n$ and then simultaneously
solve Eqs. (\ref{SELFCONS})-(\ref{EFFMASS}), together with Eq.
(\ref{IPARTICLE}).

Another interesting quantity is the nuclear symmetry energy $E_s$
defined as:
\begin{eqnarray}
E_s=\frac{1}{2} \left. \frac{\partial^2 E}{\partial \chi^2}
\right|_{\chi=0}, \nonumber
\end{eqnarray}
\noindent where $\chi=(n_n-n_p)/n$. This energy contains a purely
kinetic term $T$, and a contribution $V_s$ coming from the
isovector mesons only. The explicit expression for $E_s$ at zero
temperature, including the $a_0$ meson contribution has been
already derived, see for example \cite{KUTSCHERA2}:

\begin{eqnarray}
E_s&=&T+V_s, \label{E-S}\\
T&=&\frac{1}{12}\sum_{i=p,n} \, \frac{p_{F_i}^ 2}{E_{F_i}}, \\
V_s&=&g_r^{\ast 2} \frac{n_0}{8 m_r^2}-g_d^{\ast 2} \frac{n_0
\sum_i \,\left( \frac{M_i^\ast}{E_{F_i}}\right)^2} {4
(m_d^2+g_d^{\ast 2} A)},\label{E-S3}
\end{eqnarray}
\noindent where $p_{F_i}$ is the Fermi momentum for the i-type
particle, $E_{F_i}=\sqrt{p_{F_i}^2+M_i^{\ast 2}}$, and

\begin{eqnarray}
A=\sum_{i=p,n} \frac{3}{\pi^2 E_{F_i}}\left[M_i^{\ast 2}
p_{F_i}+\frac{p_{F_i}^3}{3}- M_i^{\ast 2} \ln\left(
\frac{p_{F_i}+E_{F_i}}{M_i^\ast}\right)\right] \nonumber
\end{eqnarray}

\section{Bulk properties of symmetric nuclear matter at T=0}
\label{SEC3}

\begin{figure}
\hspace{3.cm}
\psfig{file=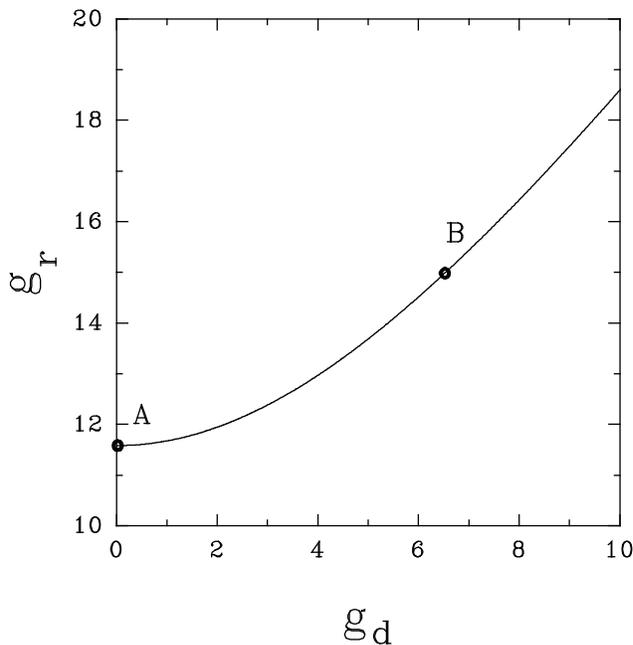,height=8.5cm} 
\caption{The relationship between the couplings $g_r$ and $g_d$,
constrained to reproduce the symmetry energy $E_s=32$ MeV at zero
temperature and at the saturation density $n_0$. The pair of couplings
$A$ and $B$ used in our calculations are marked with circles.}
\label{FIGgr-gd}
\end{figure}

In the previous section we have presented the model, which contains
several free parameters. The masses of the $a_0, \omega$ and $\rho$
mesons are taken at their physical values $m_d=984$ MeV, $m_w=783$ MeV,
and $m_r=770$ MeV respectively. We adopt the accepted value for the
$\sigma$ meson mass $m_s=550$ MeV. There remains to determine the four
coupling constants. We adjust them to reproduce the main bulk
properties of symmetric nuclear matter: the saturation density
$n_0=0.15 fm^{-3}$, the binding energy $\varepsilon_B=-15$ MeV, and the
symmetry energy $E_s=32$ MeV at zero temperature and at normal density.
Another quantity of physical interest is the isothermal compressibility
$\kappa_T$, however the DSCM provides very good values for $\kappa_T$
without imposing any further condition. Therefore we have three
physical conditions to fix the four coupling constants. Two of them,
$g_s$ and $g_w$ are univocally determined to take the values
$g_s=12.379$, $g_w=14.624$, whereas $g_r$ and $g_d$ are functionally
dependent through Eqs. (\ref{E-S})-(\ref{E-S3}) evaluated at $n=n_0$.
For our calculations we have selected two sets of couplings
$(g_r,g_d)$, denoted by $A$ and $B$: set $A=(11.583,0)$ and set
$B=(15,6.538)$, which are shown in Fig. \ref{FIGgr-gd}.

\begin{figure}[b]
\hspace{3.cm}
\psfig{file=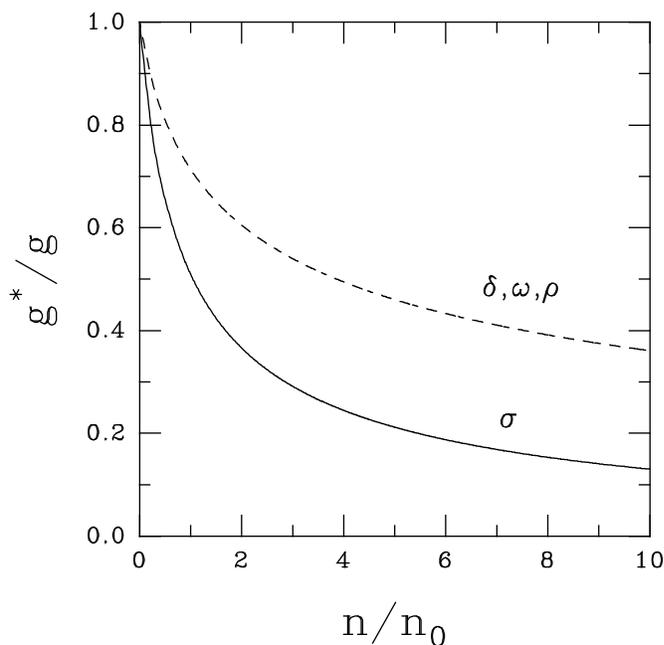,height=8.5cm} 
 \caption{The effective couplings relative to their vacuum values
in symmetric nuclear matter at T=0, in terms of the baryon density. The
solid line corresponds to the $\sigma$ channel, and the dashed line to
the $\delta, \omega$, and $\rho$ cases.} \label{FIGgS-MFA}
\end{figure}

As previously mentioned, a feature of the model proposed is the
presence of effective couplings. The behaviour of these couplings
relative to their vacuum values is the same for the $\delta, \omega,$
and $\rho$ fields, and different for the scalar $\sigma$, as discussed
in Sec. \ref{SEC2}. As can be seen in Fig. \ref{FIGgS-MFA}, the channel
corresponding to the last case is much more suppressed in dense matter.
This figure corresponds to symmetric nuclear matter, but it must be
noted that the behavior of the effective couplings depends on the
composition of the hadronic medium, {\it i.e.} they must depend on the
asymmetry coefficient $\chi$.

Although of diverse inspiration and derivation, we compare these
results with the density dependent hadron field theory (DDHFT) outcomes
\cite{LENSKE1,LENSKE2,LENSKE3,TYPEL}. For this purpose we use the
interpolating algebraic function given in \cite{LENSKE3}. Differences
are appreciable at medium and high densities. The couplings for
$\sigma, \omega,$ and $\rho$ are monotonous decreasing in both
formalisms. A dropping of $20\%$ for the isoscalar mesons, and of
$42\%$ for the $\rho$ meson is detected in the DDHFT at $n/n_0=2$. Our
results provides for the same conditions a stronger decay of $60\%$ for
the $\sigma$ coupling and of $40\%$ in the remaining cases.

\begin{figure}[t]
\psfig{file=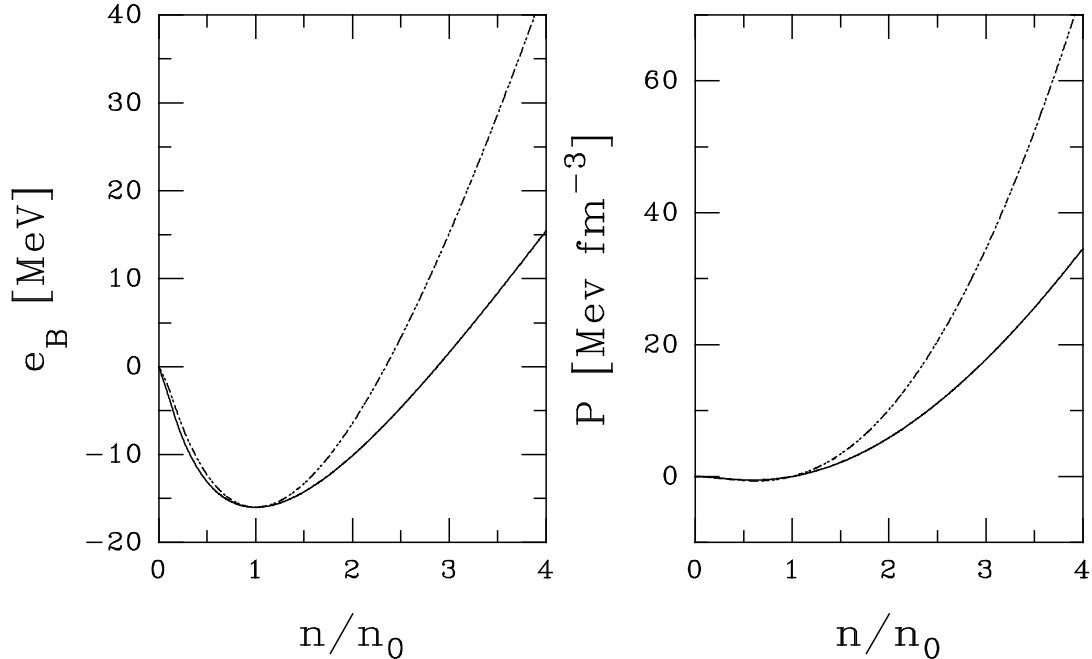,width=0.9\textwidth} 
\caption{The binding energy (left) and the pressure (right) as
functions of the baryon number density at T=0, in the MFA. Solid and
dashed lines correspond to our results and the standard DSCM
calculations, respectively.} \label{FIGEoST=0x=0}
\end{figure}

With the sets of parameters $A$ and $B$ we have evaluated some
properties of symmetric nuclear matter at zero temperature. In Fig.
\ref{FIGEoST=0x=0} we compare our results for the binding energy and
the pressure as functions of the baryon number density, with the
corresponding outcome of the standard DSCM. It can be seen that there
are not appreciable differences below $n=1.5\,n_0$, from here on both
$\varepsilon_B$ and $P$ grow more slowly in our calculations. The
isothermal compressibility is a measure of the stiffness of the
pressure, we get at the saturation density $\kappa_T=165$ MeV, against
$\kappa_T=220$ MeV for the DSCM. The lower slope of the binding energy
in our results is essentially due to the weakening of the repulsion at
higher densities induced by the normalization factor $N$. On the other
hand, the relative difference between the mean values $\bar{\sigma}$
and $\bar{\omega}$ increases with $n$ in our model, meanwhile in the
standard DSCM it approaches to zero. This gives rise to the relative
lessening of the pressure at high densities in our results.

\begin{figure}
\hspace{3cm}
\psfig{file=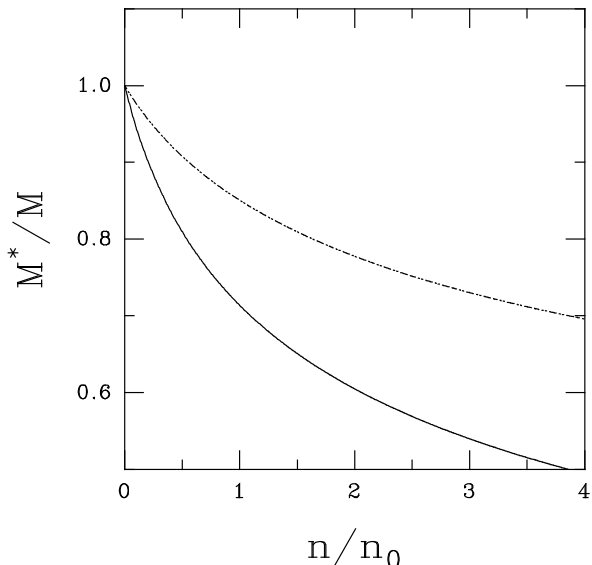,height=7.5cm}
\caption{The effective nucleon mass at zero temperature in symmetric
nuclear matter. The line convention is the same as in Fig.
\protect\ref{FIGEoST=0x=0} } \label{FIGMASS-MFA}
\end{figure}

The medium effects on the effective nucleon mass can be seen in Fig.
\ref{FIGMASS-MFA}, where a comparison with the DSCM result is made. In
both cases $M^\ast$ is positive definite and monotonous decreasing, but
the rate of falling at densities $0<n<2\,n_0$ is more pronounced in our
case because of the higher value of $g_s$ needed to reproduce the
normal properties of nuclear matter. At higher values of the baryonic
density $M^\ast$ stabilizes, due to the dynamical screening of the
effective coupling.

\begin{figure}
\psfig{file=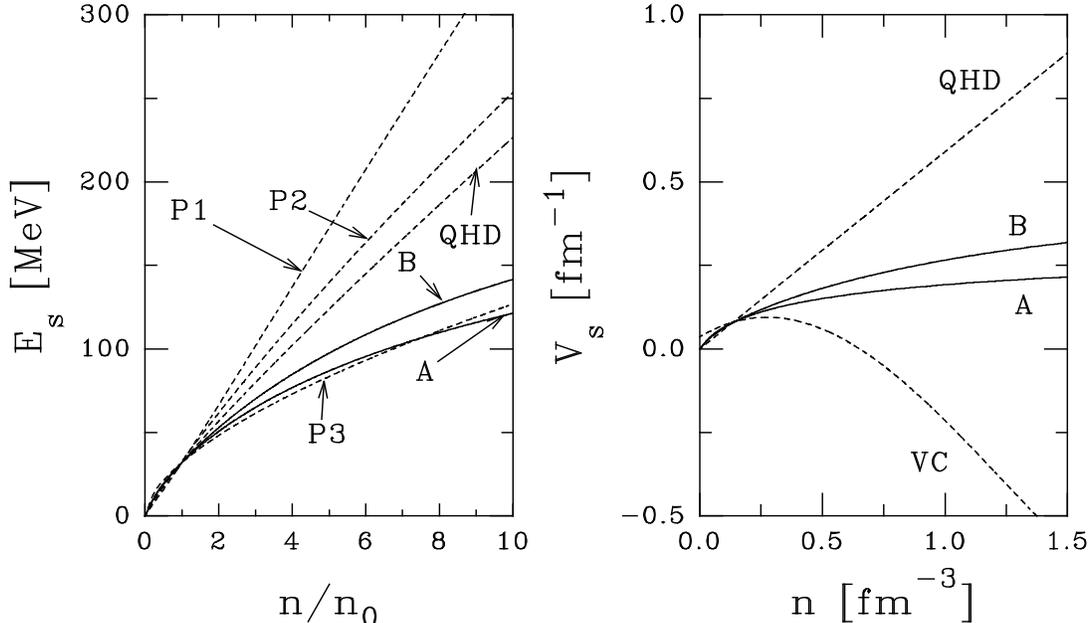,width=0.9\textwidth} 
\caption{The full symmetry energy (left panel) and the interaction
contribution $V_s$ (right panel), in terms of the baryon number
density. In both cases solid lines correspond to results with coupling
sets A and B. Dashed lines are used in the left panel for the QHD-I
model (without $\delta$ contribution), and the parameterizations P1, P2
and P3 as explained in the text. In the right panel dashed lines
correspond to the QHD-I model and to the parameterization given in Ref.
\protect\cite{KUTSCHERA3} for variational calculations (VC). }
\label{FIGESymm}
\end{figure}

 As the next step we investigate the density dependence of the
symmetry energy $E_s$. It has been profusely studied in the past, using
non-relativistic potentials as well as relativistic formulations
\cite{LI1,LI2,KUTSCHERA,KUTSCHERA1,KUTSCHERA2,KUTSCHERA3,BROWN}.
Recently $E_s$ has received attention by its applications to the study
of the structure of nuclei with a large neutron excess, produced in the
radioactive ion beam facilities. It is also a relevant subject in the
evolution of neutron stars, determining the composition of the ground
state and the cooling mechanism \cite{LATTIMER}, or the phase
transition to quark matter \cite{KUTSCHERA3}. Furthermore, it has been
proposed that the ratio of neutrons to protons in the pre-equilibrium
stage of collisions between neutron rich nuclei could distinguish the
asymmetric contribution of the nuclear equation of state \cite{LI1}.

Different theoretical predictions for $E_s$ produce rather dissimilar
density dependences. In the left part of Fig. \ref{FIGESymm} we compare
our results for the symmetry energy coefficient, with and without the
contribution of the $a_0$-meson,  with other commonly used
descriptions. We include the result from the QHD-I model \cite{WALECKA}
without scalar-isovector coupling, and other three cases labeled P1,
P2, and P3. The latter correspond to the phenomenological
parameterizations \cite{LATTIMER2}:

\begin{eqnarray*}
E_s&=&\frac{3}{5} \left( 2^{2/3}-1\right)e_F
\left[u^{2/3}-F(u)]+E_{s 0} F(u)\right],
\end{eqnarray*}

\noindent where $u=n/n_0$, $e_F$ is the non-relativistic Fermi energy
at the saturation density, and the function $F(u)$ takes the forms
$F_1(u)=2 u^2/(1+u)$, $F_2(u)=u$, and $F_3(u)=\sqrt{u}$ for the curves
denoted as P1, P2 and P3, respectively. All the curves are almost
coincident for densities $n<1.5\, n_0$, but their mutual differences
become significant for densities above that limit. The exception
corresponds to the cases A and P3, which differ each other only by
negligible amounts in all the range of shown densities. From Eq.
(\ref{E-S3}) it can be seen that the contributions to $V_s$ of the
iso-vector $\delta$ and $\rho$ mesons are opposite in sign. However
choosing $g_d\neq 0$ brings on an enhanced behaviour of $E_s$, because
the value of $g_r$ required to adjust $E_s=32$ MeV at $n=n_0$ is bigger
than $g_d$ (see Fig. \ref{FIGgr-gd}).
 It must be noted that the rate of growth of the cases A,
B, and P3 decreases with density, whereas it remains approximately
constant for the curves P1, P2 and QHD-I.

The effect of polynomial self-interactions of the $\sigma$ field in QHD
models has been studied in \cite{GRECO}. Both, the inclusion of
exchange terms and of the $\delta$ coupling in Hartree approximation
enhance the density dependence of $E_s$, and therefore diverge from
curves A and B in the left part of Fig. \ref{FIGESymm}, being more
alike to the P1 parametrization.

The behavior of $E_s$ depends on the method of evaluation and the model
of interaction used, the latter defines the $V_s$ term. In the right
part of Fig. \ref{FIGESymm} we display the interaction contributions
for $E_s$ obtained with the sets A and B, and we compare them with
$V_s$ extracted from the QHD-I model \cite{WALECKA} and with the
parameterization given by \cite{KUTSCHERA3} for the variational
calculations (VC) made in \cite{WIRINGA}. It can be seen that our
results are intermediate between QHD-I and the VC results. A
characteristic behaviour of the VC is that $V_s$ becomes negative for
densities bigger than certain typical value, causing the disappearance
of protons in neutron stars at high densities.

\noindent From the behavior of the symmetry terms shown in Fig.
\ref{FIGESymm}, we expect that the fraction of protons in star matter
should be lower in our results as compared, for instance, with the
QHD-I model prediction, although this fraction remains non vanishing
for all densities in our case. The inclusion of the $\delta$ coupling
(curve B) slightly increases the presence of protons.

\section{The equation of state of asymmetric nuclear matter}
\label{SEC4}

We study here the properties of nuclear matter at finite temperature by
taking the asymmetry coefficient $\chi$ as a free parameter. In the
next section the isospin asymmetry will be determined by the conditions
of electric charge neutrality and matter stability against electroweak
decay.

\begin{figure}
\hspace{3cm}
\psfig{file=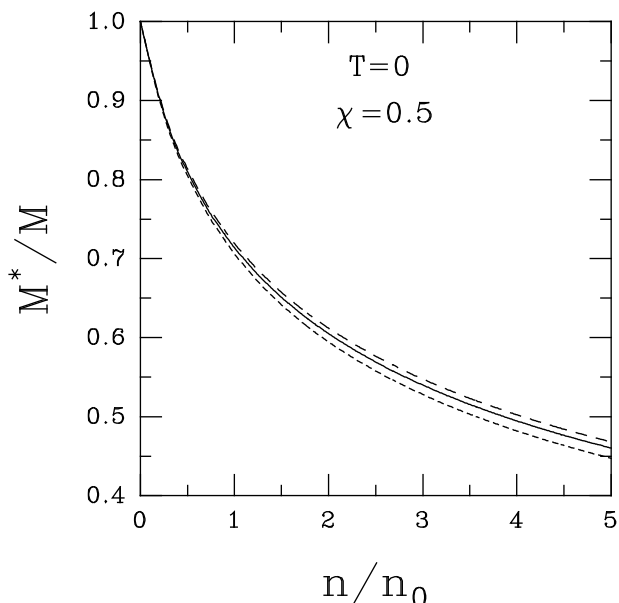,height=8cm} 
\caption{The effective masses of proton and neutron as functions of the
relative baryon number density for asymmetric nuclear matter at zero
temperature. The solid (dashed) line corresponds to results with the
coupling set A (B).} \label{FIGMASSY}
\end{figure}

In first place we inspect the density dependence of the nucleon
effective mass for fixed $\chi$. In Fig. \ref{FIGMASSY} we compare
results with and without $\delta$ coupling at T=0 and $\chi=0.5$.
 For $g_d=0$ (set A) proton and neutron masses are degenerate, and for
$g_d \neq 0$ (set B) the neutron (proton) mass is lowered
(enhanced) due to medium effects. The splitting is heightened as
the density increases.

Temperature effects are minimal in the range $0< T < 100$ MeV, and more
noticeable at high densities. For example, when the coupling set A is
chosen, an increment of about 5 MeV in the nucleon effective mass is
observed at $n=5 n_0$ as the temperature is raised from $T=0$ to
$T=100$ MeV, at a given $\chi$. Of the same magnitude but opposite in
sign is the effect of increasing the asymmetry from $\chi=0$ to
$\chi=1$ at a fixed temperature. When the coupling set B is used, it is
found that the in-medium mass splitting $\Delta M^\ast=M^\ast_p -
M^\ast_n$ decreases when the temperature is raised at fixed $\chi$. On
the other hand, $\Delta M^\ast$ is enhanced when the asymmetry is
isothermally increased. Numerical values of this mass splitting depend
on the set of couplings used, and we estimate the magnitude of both
temperature and asymmetry effects calculating $\Delta M^\ast$ with the
set B at $n=5\, n_0$. In this case the splitting reduces about 5 MeV in
neutron matter when the temperature covers the range $0< T < 100$ MeV,
but an increment of approximately 50 MeV is found in $\Delta M^\ast$ if
$\chi$ is varied between $\chi=0$ and $\chi=1$ at fixed temperature.

\begin{figure}
\hspace{3cm}
\psfig{file=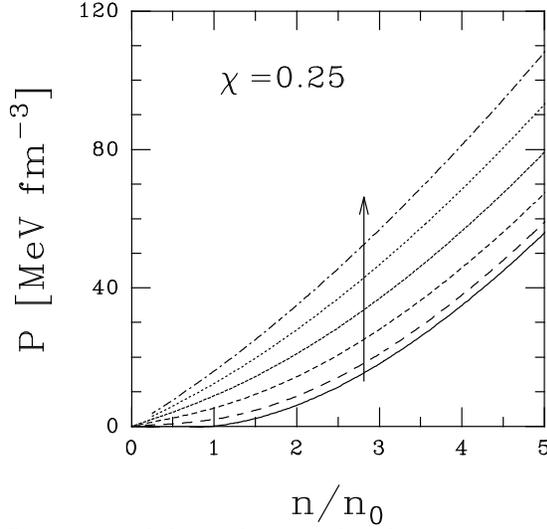,height=7cm} 
\caption{The pressure P in terms of the nucleon number density at
$\chi=0.25$. The different curves correspond to temperatures $T=0, 20,
40, 60, 80,$ and $100$ MeV. The arrow indicates the sense of growing
temperatures. The results shown are obtained with the set A.}
\label{FIGPVSN-TX0}
\end{figure}

\begin{figure}
\psfig{file=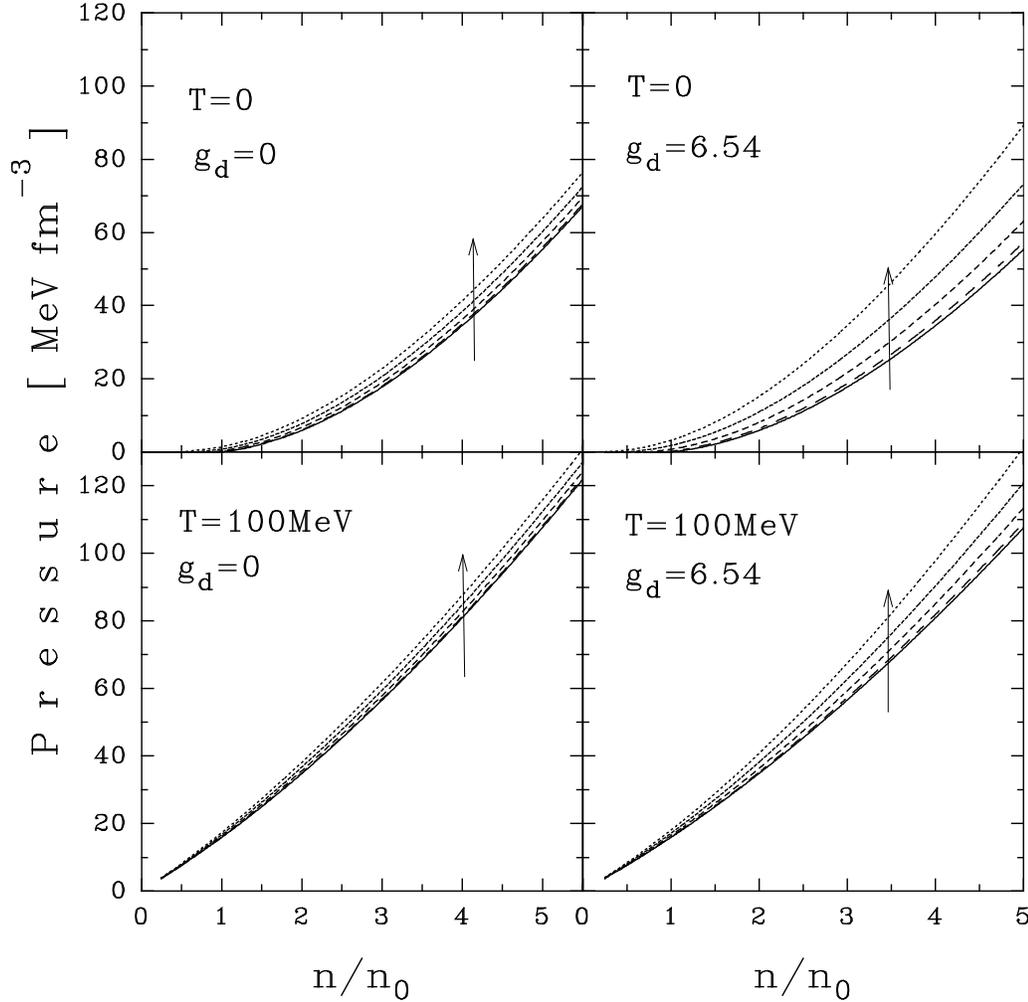,width=0.85\textwidth} 
\caption{The pressure as function of the nucleon number density.
Different types of lines correspond to the asymmetry coefficients
$\chi=0, 0.25, 0.50, 0.75,$ and $1.0$, respectively. The arrows
indicate the direction of increasing $\chi$. The top (bottom)
panel represents calculations for T=0 (T=100 MeV), and the left
(right) side is devoted to results using the coupling set A (B).}
\label{FIGPVSN-XT0}
\end{figure}

The thermodynamical pressure $P$ has been evaluated using Eq.
(\ref{PRESSURE}), for several temperatures $0 < T < 100$ MeV, and
several asymmetries $0 < \chi < 1$. As expected, increasing the
temperature produces an enhancement of the pressure. This effect is
strengthened by raising the asymmetry. The quantitative behavior of the
pressure can be seen in Figs. \ref{FIGPVSN-TX0} and \ref{FIGPVSN-XT0}.
 In the first one we plot
the pressure as a function of the number density at fixed asymmetry
$\chi=0.25$ and for several temperatures.
 For $T\geq 20$ MeV it is a monotonous increasing
function of the density, whereas for $T=0$ it exhibits a region of
instability for densities below $n_0$. This instability gives rise to a
liquid-gas phase transition \cite{WALECKA}. The results in Fig.
\ref{FIGPVSN-TX0} correspond to the set A. By using the set B
qualitatively similar results are obtained.

The relevance of the asymmetry in our calculations can be observed in
Fig. \ref{FIGPVSN-XT0}. The higher the values of the asymmetry the
stiffer the pressure raises, this effect being emphasized when the
coupling $g_d$ is non zero. The liquid-gas instability remains for low
$T$ and $n$, disappearing for $\chi$ close to $1$.

\section{Hadronic matter in $\beta$-equilibrium.}
\label{SEC5}

The conditions in the interior of certain stellar objects like
protoneutron stars, requires additional degrees of freedom to be
included in the lagrangian density of Sec. \ref{SEC2}. Due to the large
densities reached in such systems, several physical phenomena could
take place. The appearance of mesons and baryons with strangeness, pion
and/or kaon condensation, the chiral symmetry restoration, and the
phase transition to a quark-gluon plasma are some of the expected
processes. They must be taken into account, in order to properly
describe the high density behavior of the equation of state. In this
section we complete the model proposed by including the hyperons
$\Lambda, \Sigma,$ and $\Xi$, but we do not treat explicitly the chiral
symmetry and quark degrees of freedom. Therefore our results should be
valid until fluctuations preceding any phase transition become
relevant. However we present here calculations in the range $0< n/n_0 <
10$ for the sake of comparison.

 As anticipated in Sec. \ref{SEC2}, the modifications in the
lagrangian density are straightforward since we retain the form of the
interaction for all the baryons. A sum over the full octet $N, \Lambda,
\Sigma,$ and $\Xi$ must be considered in Eq. (\ref{DSCM}), and new
couplings $g_s, g_d, g_w,$ and $g_r$ are introduced for the hyperons.
Furthermore, the vertex between the $\rho$ meson and the baryon B must
be modified by including an appropriate coefficient: $I_{B3}=1/2$ for
proton and $\Xi^0$, $I_{B3}=-1/2$ for neutron and $\Xi^-$, $I_{B3}=1$
for $\Sigma^+$, $I_{B3}=0$ for $\Lambda$ and $\Sigma^0$, and
$I_{B3}=-1$ for $\Sigma^-$.

The new couplings should be fixed to reproduce some relevant quantity,
according to the phenomenological approach. We proceed in this way to
determine the $\sigma$- and $\omega$-$\Lambda$ couplings. Using
hypernuclei data the $\Lambda$ binding energy can be extrapolated to be
$\varepsilon_\Lambda=-28$ MeV at $n_0$, thus we obtain $g_{s
\Lambda}=2.335$, $g_{w \Lambda}=2.099$.
 For the other hyperons there are not accurate
experimental data. Different arguments are commonly used to get
numerical values, like $SU(6)$ symmetry or vector meson dominance. For
simplicity and to carry out computations, we adopt $g_{s,w
\,\Sigma}=g_{s,w \,\Xi}=g_{s,w \, \Lambda}$ and $g_{r,d
\,\Sigma}=g_{r,d \,\Xi}=g_{r,d \, \Lambda}=g_{r,d}$, without any
further justification. With this choice we obtain at $n_0$ very similar
binding energies: $\varepsilon_\Sigma=-28.11$ MeV, and
$\varepsilon_\Xi=-28.27$ MeV for the $\Sigma$ and $\Xi$ hyperons.

Neutron star matter is electrically neutral, by additional
contributions coming from electrons and muons. Leptons are included by
means of Dirac free particle terms in the lagrangian of Eq.
(\ref{DSCM}). The equilibrium for $\beta$ decay imposes constraints
among the baryon and lepton chemical potentials: $\mu_B=\mu_n - q_B
\mu_e$. Here we have used $q_B$ for the baryon electric charge in units
of the positron charge, $\mu_n, \mu_e,$ and $\mu_B$ represents the
chemical potentials for neutron, electron and the baryon B,
respectively. On the other hand, the electric
 charge neutrality imposes $0=-\sum_l n_l+ \sum_B q_B
n_B$, with $n_l$ and $n_B$ indicating the number density for leptons
and baryons. At zero temperature we consider the Fermi momentum $p_F$,
writing: $n_i=p_{F\, i}^3/(3 \pi^2)$, $\mu_l=\sqrt{p_{F\,l}^2+m_l^2}$
for leptons, and $\mu_B=\sqrt{p_{F\,B}^2+M_B^{\ast 2}}+ g_{w \,
B}^{\ast} \bar{\omega}+ g_r^\ast I_{B3} \bar{\rho}$ for baryons. The
efective mass $M^\ast_B$ is a generalization of eq. (\ref{EFFMASS}),
$M^\ast_B=(M_B-I_B g_d \bar{\delta})/N_B$ with $I_B=1$ for $p,
\Sigma^+$, and $\Xi^0$, $I_B=0$ for $\Lambda$ and $\Sigma^0$, and
$I_B=-1$ for $n, \Sigma^-$, and $\Xi^-$. We have used $N_B=1+ g_{sB}
\bar{\sigma} / M_B$, where $M_B$ is the averaged mass of the baryon
isomultiplet B.

\begin{figure}
\psfig{file=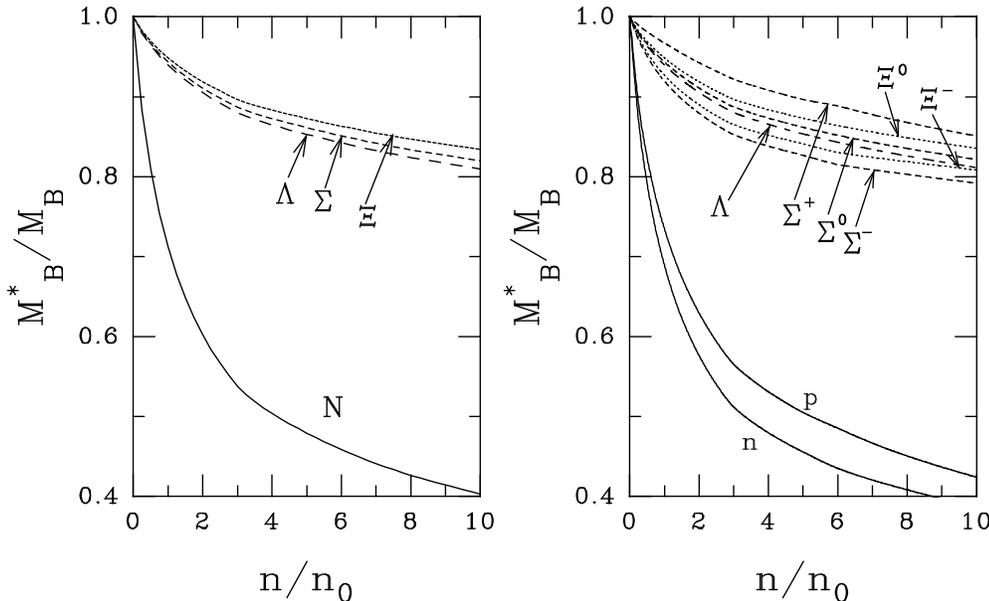,height=8cm} 
\caption{The effective baryon masses in terms of the total baryon
number density. The left (right) panel corresponds to calculations with
set A (B). Different type of lines are assigned to each iso-multiplet,
as indicated in both panels.} \label{FIGHYPMASS}
\end{figure}

The effective baryon masses as functions of the baryonic number density
are shown in Fig. \ref{FIGHYPMASS}. In the results corresponding to set
A, each isomultiplet remains degenerate in mass. The variation of the
hyperon masses are much more moderate than for the nucleon masses. As a
consequence of the specific interaction used, the heavier the baryon
considered, the weaker the density dependence of its effective mass is.
Using the coupling set B the isospin degeneracy is removed, enhancing
or dropping the mass of particles with positive or negative isospin
projection, respectively, as it is depicted on the right part of Fig.
\ref{FIGHYPMASS}.

\begin{figure}
\psfig{file=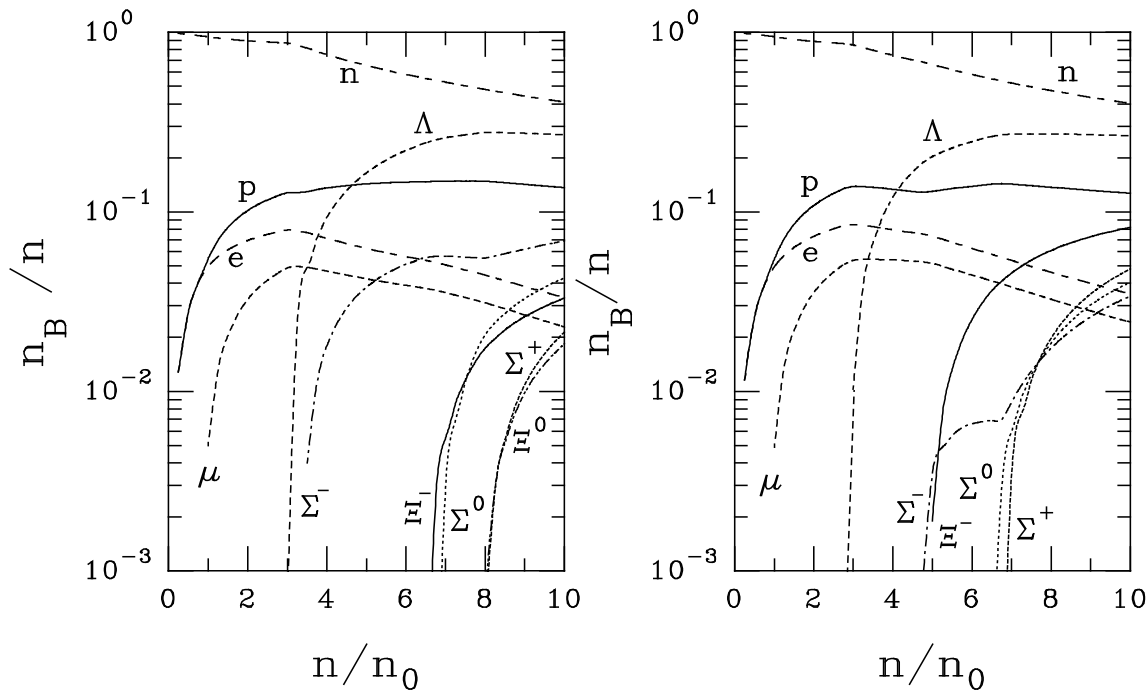,width=0.95\textwidth} \vspace{3mm} \caption{The
fraction of leptons and baryon species present in hadronic matter in
$\beta$-equilibrium at $T=0$. The different curves are labeled with the
corresponding particle name. The left (right) panel corresponds to
calculations with the coupling set A (B). } \label{FIGHYP-POP}
\end{figure}

In Fig. \ref{FIGHYP-POP}  the relative population of the baryonic
species is shown in terms of the total particle number, at $T=0$. In
the range of densities studied the full baryon octet is present, with
exception of the $\Xi^0$ when the coupling set B is used. The results
obtained with $g_d=0$ and $g_d=6.538$ are very similar for the leptons
and the lightest baryons (p, n, and $\Lambda$). Differences between
them become noticeable in the growth of populations of the heavier
fermions $\Sigma$ and $\Xi$. The more obvious is the early appearance,
at $n \simeq 3.3\, n_0$, and predominance of $\Sigma^-$ particles in
the results with the coupling set A.

The equilibrium baryonic population is not perturbed by the
presence of the $\delta$ meson (set B) at low and medium
densities. The effect of turning on the $\delta$ interaction,
is twofold and is emphasized at high densities. In first place the
baryon-$\delta$ interaction enhances the effective mass of $\Xi^0$ and
diminishes that of $\Sigma^-$ and $\Xi^-$, increasing and lowering the
corresponding thresholds. The more evident consequence of this is the
absence of $\Xi^0$ particles in the range $0<n/n_0<10$ (right panel of
Fig. \ref{FIGHYP-POP}). In second place the coupling $g_r$ grows with
$g_d$, affecting more strongly to the iso-triplet ${\bm \Sigma}$ than
the iso-duplet ${\bm \Xi}$, due to the factor $I_{B3}$. Since the
$\rho$ meson contribution to the chemical potential is positive and
greater for $\Sigma^-$ than for $\Xi^-$, this causes the appearance of
$\Sigma^-$ and of $\Xi^-$ to be delayed and anticipated, respectively,
going from the left to the right panel of Fig. \ref{FIGHYP-POP}. The
$\Lambda$ and $\Sigma^0$ baryons, which do not couple to the isovector
mesons, do not show appreciable changes in their distributions. On the
other hand, from the comparatively earlier raising of the $\Sigma^+$
and  $\Xi^-$ population obtained with set B, it is possible to infer
that, in absolute values, the  $\delta$ contribution to the baryonic
chemical potentials lies between one half and the total $\rho$
contribution. Of course, these results are partially a consequence of
our assumption of equal couplings $g_d$ and $g_r$ for all the hyperons
considered.

\begin{figure}
\hspace{3cm}
\psfig{file=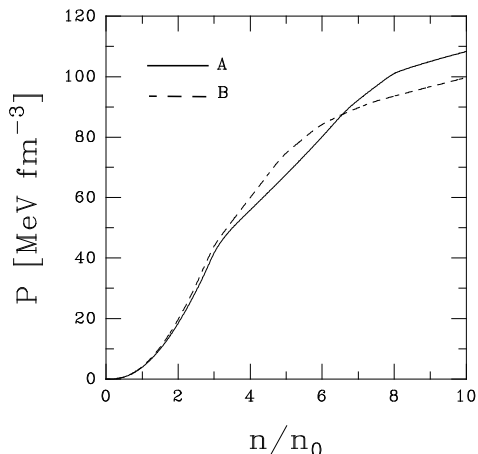,height=6cm} 
\caption{The pressure of hadronic matter in $\beta$-equilibrium at
$T=0$. As indicated in the figure, solid (dashed) line corresponds to
calculations using the coupling set A (B).} \label{FIGHYPress}
\end{figure}

The pressure in terms of the baryon number density is exhibited in Fig.
\ref{FIGHYPress}. There are abrupt changes of slope in the curve
corresponding to the set A, which coincide with the appearance of
hyperons. Similar changes, but more attenuated, take place in the curve
with the set B.

\section{Meson propagation in asymmetric nuclear matter}
\label{SEC6}

Medium effects in the meson properties have received attention in the
later years, as they could carry the signals of phase transitions in
the hadronic environment. As previously stressed, we expect our results
to be valid out of the vicinity of the transition point.

 In the MFA mesons are treated as classical
fields, with constant mass. In order to include quantum corrections we
must go beyond the MFA. This can be done in the relativistic random
phase approximation (RRPA), using the linearized residual interaction
of Eq. (\ref{Lres3}). In this approach the meson propagators are
corrected by incorporating the baryon bubble diagrams at all orders, by
using the Dyson-Schwinger equation. From the corrected propagator the
effective meson mass can be extracted. This procedure has been applied
in QHD calculations, see for example \cite{WALECKA} and references
listed therein. Specific computations with the DSCM can be found in
\cite{BHATTAC,AGUIRRE}.

\begin{figure}
\hspace{3.5cm} \psfig{file=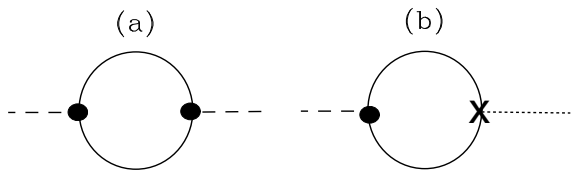,width=0.47\textwidth} \vspace{3mm}
\caption{Feynman diagrams included in the RRPA. Case a (b) corresponds
to pure (mixing) meson propagation. The solid line stands for baryon
propagator, dashed and dotted lines represent meson propagators of
different types, and the filled circle and the cross their respective
vertices.} \label{FIGDIAG}
\end{figure}

At second order, the one-loop proper polarization insertions comprise
the diagrams shown in Fig. \ref{FIGDIAG}. The case (a) represent the
propagation of a pure meson field, and the case (b) the mixing
amplitude of different mesonic types. Due to baryon current
conservation, the proper polarization for all the mesons can be written
in terms of a few components. Some of them are divergent and requires
an appropriate regularization. For this purpose, we follow the scheme
outlined in \cite{AGUIRRE}. We do not deduce those results here, but we
reproduce the main equations for the sake of completeness.

The formalism is best described within a generalized meson
propagator, in a matrix representation of dimension equal to the
sum of the mesonic degrees of freedom. For example the generalized
free meson propagator ${\cal P}^0$, has 
in its diagonal blocks the free meson propagators $S^0(q)$, $D^{0\,A
B}(q)$, $W_{\mu \nu}^0(q)$, and $R_{\mu \nu}^{0\,AB}(q),$  for the
$\sigma (x), \delta (x), \omega (x),$ and $\rho (x)$  fields,
respectively, and null matrices in the complementary spaces:
\begin{eqnarray}{\cal P}^0_{\alpha \beta}(q)&=&
\left(\begin{array}{rrrr}
S^0(q)\stackrel{:}{:}&&&\\
..........&.........&&\\
\stackrel{:}{:}&D^0(q)\stackrel{:}{:}&&\\
&..........&.........&\\
&\stackrel{:}{:}&W^0(q)\stackrel{:}{:}&\\
&&...........&.........\\
 &&\stackrel{:}{:}&R^0(q)
\end{array}\right)_{\alpha \beta}. \nonumber
\end{eqnarray}

A similar expression holds for the full generalized propagator
${\cal P}$, but the complementary spaces are filled with the
mixing meson propagators:

\begin{eqnarray}{\cal P}_{\alpha \beta}(q)&=&
\left(\begin{array}{cccc}
S(q)&M_{\sigma \delta}(q)&M_{\sigma \omega}(q)&M_{\sigma \rho}(q)\\
M_{\delta \sigma}(q)&D(q)&M_{\delta \omega}(q)&M_{\delta \rho}(q)\\
M_{\omega \sigma}(q)&M_{\omega \delta}(q)&W(q)&M_{\omega \rho}(q)\\
M_{\rho \sigma}(q) &M_{\rho \delta}(q)&M_{\rho \omega}(q)&R(q)
\end{array}\right)_{\alpha \beta}. \nonumber
\end{eqnarray}

The Dyson-Schwinger equation can be used to solve for ${\cal
P}^{-1}(q)$:
\begin{eqnarray}
{\cal P}^{-1}_{\alpha \beta}(q)&=&{\cal P}^{0 \,-1}_{\alpha
\beta}(q)- \Pi_{\alpha \beta}(q), \nonumber
\end{eqnarray}
\noindent
 where we have introduced the generalized polarization
insertion

\begin{eqnarray}\Pi_{\alpha \beta}(q)&=&
\left(\begin{array}{cccc}
\Pi_s(q)&\Pi_{\sigma \delta}(q)&\Pi_{\sigma \omega}(q)&\Pi_{\sigma \rho}(q)\\
\Pi_{\delta \sigma}(q)&\Pi_d(q)&\Pi_{\delta \omega}(q)&\Pi_{\delta \rho}(q)\\
\Pi_{\omega \sigma}(q)&\Pi_{\omega \delta}(q)&\Pi_w(q)&M_{\omega \rho}(q)\\
\Pi_{\rho \sigma}(q) &\Pi_{\rho \delta}(q)&\Pi_{\rho
\omega}(q)&\Pi_r(q)
\end{array}\right)_{\alpha \beta}. \nonumber
\end{eqnarray}

Since we are
primarily interested in the propagation of the pure meson fields,
we do not consider mixing polarizations. The formulae for the
one-loop diagonal components are as follows:

\begin{eqnarray}
i \Pi_s(q)&=&\sum_{B}g_{sB}^{\ast \,2}\int \frac{d^4p}{(2 \pi)^4}
\left\{ Tr\left[ G_B(q)G_B(q+p)\right] \frac{\vspace{1.cm}}{}+
\sum_{\lambda}\frac{2\bar{\phi}_\lambda}{M_B}Tr\left[
G_B(q)\Gamma_\lambda G_B(q+p)\right] \right. \nonumber \\ &+&\left.
\sum_{\lambda,\lambda^\prime} \bar{\phi}_\lambda
\bar{\phi}_{\lambda^\prime} Tr\left[ G_B(q)\Gamma_\lambda
G_B(q+p)\Gamma_{\lambda^\prime}\right]/M_B^{2} \right\},\nonumber
\\ i\Pi_d^{AC}(q)&=&g_d^{\ast \, 2}\sum_B \int \frac{d^4p}{(2
\pi)^4} Tr\left[ G_B(q)T^A G_B(q+p)T^C \right], \nonumber \\ i
\Pi_w^{\mu \nu}(q)&=&\sum_B g_{wB}^{\ast \, 2} \int \frac{d^4p}{(2
\pi)^4} Tr\left[ G_B(q)\gamma^\mu G_B(q+p)\gamma^\nu \right], \nonumber
\\ i \Pi_{r \, \mu \nu}^{AC}(q)&=&g_r^{\ast \, 2}\sum_B \int
\frac{d^4p}{(2 \pi)^4} Tr\left[ G_B(q)\gamma_\mu T^A G_B(q+p)\gamma^\nu
T^C \right], \nonumber
\end{eqnarray}
\noindent where  the index $B$ runs over all the baryons considered,
and $\lambda, \lambda^\prime$ in the first equation runs over the meson
fields $\delta, \omega,$ and $\rho$. The vertices $\Gamma_\lambda$ have
been described in Sec. \ref{SEC2}. The baryon propagators $G_B(q)$ are
evaluated in the MFA. In our calculations we only need the transversal
component in the Lorentz indices, and the third component of isospin.
For this purpose we use $T^3=\tau^3$ for the nucleon and $\Xi$, $T^3=1$
for the $\Lambda$, and $T^3=$diag$(1,0,-1)$ for the $\Sigma$ particle.
The referred expressions contain particle-antiparticle, particle-hole,
and Pauli blocking contributions. The former one is divergent, to
extract finite contributions we apply the regularization scheme
outlined in \cite{AGUIRRE}. The Lorentz scalar contributions,
containing the integrand $Tr\left[ G_B(q)G_B(q+p)\right]$, remains
undefined by a constant $\lambda$, related to the covariant derivative
of the polarization evaluated at the regularization point. We take this
constant as a free parameter to analyze the possible dynamical regimes.
There are two independent parameters $\lambda_s$ and $\lambda_d$
corresponding to the diagonal components $\Pi_s$ and $\Pi_d$
respectively. We require null contribution for the polarization
evaluated on the meson mass shell, at zero baryon density and
temperature. Thus we obtain for the finite particle-antiparticle
contribution of the baryon-B bubble:

\begin{eqnarray}
\Pi_{v B}^{\prime \,00}(q) &=&\frac{g_{vB}^{\ast \,2}}{2
\pi^2}q^2\int_0^1 dz \, z(1-z)\,\,\ln\left[ \frac{M_B^{\ast
2}-z(1-z){\sf q}^2} {M_B^2-z(1-z)m_v^2}\right], \nonumber \\ \Pi_{v B
}^{\prime \,33}(q)&=&\frac{{\sf q}^2}{q^2}\,\, \Pi_{v B}^{\prime
\,00}(q), \nonumber
\\ \Pi_{c B}^{\prime}(q)&=& \lambda_c \frac{g_{cB}^{\ast 2}}{8 \pi^2}(m_B^{\ast
2}m_c^2- {\sf q}^2) -\frac{3 g_{cB}^{\ast 2}}{4 \pi^2} \int_0^1 dz
\,[M_B^{\ast 2}-z(1-z){\sf q}^2] \,\, \ln\left[\frac{M_B^{\ast
2}-z(1-z){\sf q}^2} {M_B^2-z(1-z)m_c^2}\right], \label{BBuble}
\end{eqnarray}

\begin{figure}
\psfig{file=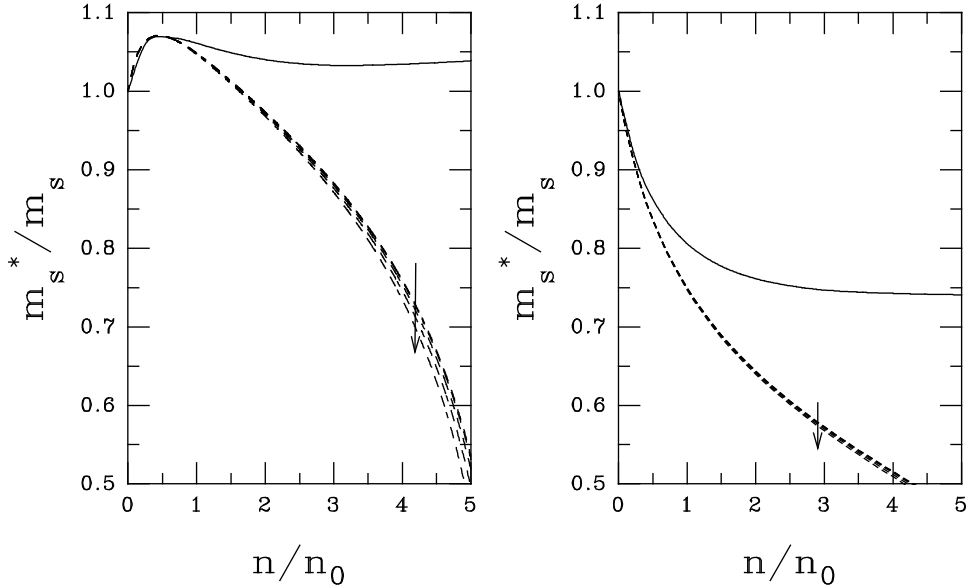,width=0.8\textwidth} 
\caption{The effective $\sigma$-meson mass as a function of the
baryonic density at T=0. In the left (right) panel the results for the
regularization parameter $\lambda_s=10$ ($\lambda_s=100$) are plotted.
In each case the solid line corresponds to hadronic matter in
$\beta$-equilibrium, and the dashed lines to nuclear matter with
asymmetry coefficients $\chi=0, 0.25, 0.50, 0.75,$ and $1.$ The arrow
indicates the direction of growing $\chi$ in the last case.}
\label{FIGMSIGMA}
\end{figure}

\noindent where the index $v=w,r$ runs over the vector mesons, and
$c=s,d$ runs over the scalar ones. In the case of  isovector
polarizations, it must be regarded as the (3,3) isospin component.
Furthermore we have used ${\sf q}^2=q_\mu q^\mu$, $q$ is the modulus of
the spatial component of the momentum, and $m_B^\ast=M_B^\ast/M_B$.

Once the polarization has been properly defined, we introduce the
effective meson masses $m_s^\ast, m_d^\ast, m_w^\ast,$ and $m_r^\ast$.
They have been defined as the zeroes of the corresponding inverse
propagators at zero vector momentum, {\em i.e.} the $p_0$ solutions of:
\begin{eqnarray}
{\cal P}^{-1}_{a a}(p_0,p=0)&=&{\cal P}^{0\, -1}_{a
a}(p_0,p=0)-\Pi_{aa}(p_0,p=0) =0, \nonumber
\end{eqnarray}
\noindent for $a=s,d,w,$ and $r$.

\begin{figure}
\psfig{file=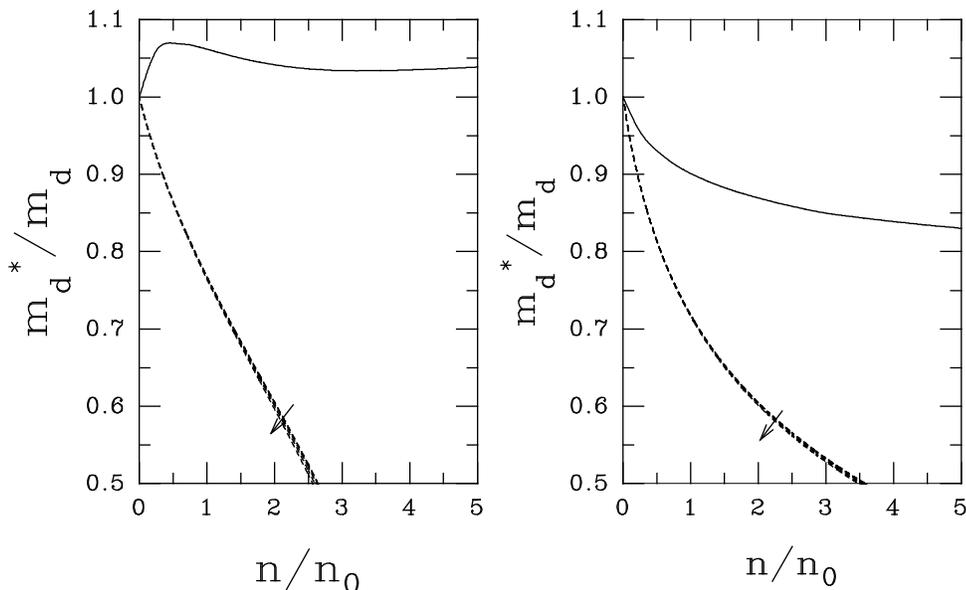,width=0.8\textwidth} 
\caption{The effective $a_0$-meson mass as a function of the density at
T=0. The left (right) panel corresponds to the values $\lambda_d=10$
($\lambda_d=100$). Solid line stands for $\beta$-stable hadronic
matter, and dashed lines correspond to asymmetric nuclear matter with
asymmetry $\chi=0.25, 0.50, 0.75,$ and $1$. The arrows indicate the
direction of growing $\chi$.} \label{FIGMDELTA}
\end{figure}

\begin{figure}
\psfig{file=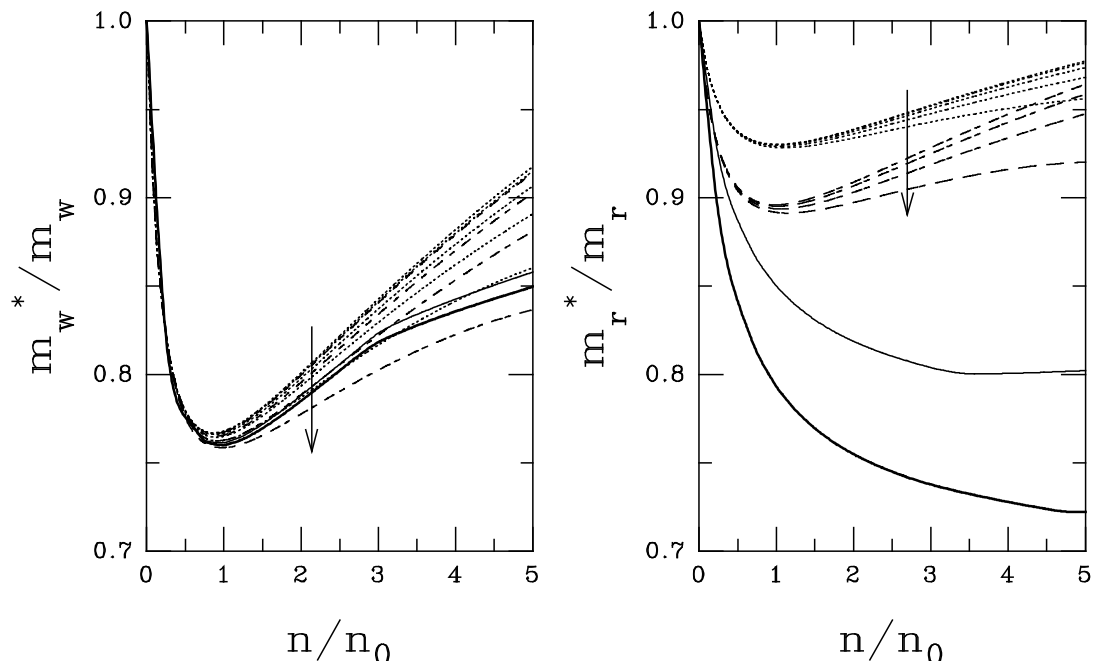,width=0.9\textwidth} 
\caption{The density dependence of the vector mesons. The left panel
corresponds to the iso-scalar $\omega$ meson, the right panel to the
iso-vector $\rho$ meson. In both cases the bold (thin) solid line
corresponds to $\beta$-stable matter with coupling set B (A), and
dashed (dotted) lines correspond to asymmetric nuclear matter with
$\chi=0, 0.25, 0.50, 0.75,$ and $1$ using the set B (A). The arrow
indicates the direction of growing $\chi$.} \label{FIGMVECTOR}
\end{figure}

In Figs. \ref{FIGMSIGMA}-\ref{FIGMVECTOR} we display the numerical
results for the density dependence of the meson masses at $T=0$, under
different compositions of the hadronic medium. In Fig. \ref{FIGMSIGMA}
the behavior of the $\sigma$ meson mass is presented. The results
correspond to the coupling set A, there are no appreciable differences
with respect to the calculations using set B .

\noindent It can be seen that in $\beta$-stable hadronic matter the
mass is almost constant at high densities. In the case of nuclear
matter at constant asymmetry the density dependence is more pronounced,
and monotonous decreasing. The asymmetry dependence is small, as this
figure shows.

Fig. \ref{FIGMDELTA} is devoted to the $a_0$ meson mass. The curves for
matter in $\beta$-equilibrium are qualitatively similar to those
corresponding to the $\sigma$ meson. In asymmetric nuclear matter its
behavior is much more striking. For $\lambda_d=10$ the mass becomes
zero at $n/n_0\simeq 4$, whereas for $\lambda_d=100$ it decreases
sharply but never vanishes in the whole range considered.

The masses for the vector mesons are shown in Fig. \ref{FIGMVECTOR}.
For the $\omega$ meson the behavior is almost independent of the
composition of the hadronic environment, sensible departures are
observed only for extreme densities. This is not the case of the $\rho$
meson mass, a clear difference among $\beta$-stable matter and
asymmetric nuclear matter is shown, even at low densities.

Since $\Pi_{aa}(p)$ receives the contribution of all the baryonic
species considered, the mesonic effective masses are strongly
influenced by the inclusion of hyperons,  even at densities close to
the normal saturation value. It must be noted that the
particle-antiparticle term coming from the baryon-B bubble contributes
even when this particle is not present on its Fermi shell. The hyperon
particle-antiparticle contributions at medium and high densities causes
the stabilization of the mesonic masses in neutral $\beta$-stable
matter. This can be appreciated in Figs.
\ref{FIGMSIGMA}-\ref{FIGMVECTOR} where they exhibit a weaker density
dependence as compared to nuclear matter results. The magnitude of this
effect is distinct for each type of meson, depending on the strength of
its coupling to the hyperons. This fact explains why the omega-meson
mass is little affected by the presence of hyperons relative to the
rho-meson mass.

In order to compare with other similar works, one can appreciate that
at high densities $m_s^\ast$ is monotonous decreasing and remains below
its vacuum value (see Fig. \ref{FIGMSIGMA}). On the contrary, the
results found in QHD-I model calculations for symmetric nuclear matter
give increasing values of it for $n\geq \, n_0$ \cite{SAITO}.
Meanwhile, the behavior of $m^\ast_w$ is more alike to ours. The
inclusion of polynomic self-interactions of the $\sigma$ and $\omega$
fields \cite{CAILLON} produces increasing or fairly constant density
dependence for $m^\ast_w$, but if the direct $\sigma$-$\omega$
interaction is omitted it results in a strongly decreasing behavior
\cite{CAILLON}. On the other hand, the $\rho$ meson mass can be
compared with the results shown in \cite{SHIOMI}. In both cases
$m^\ast_r$ is below its vacuum value, for instance at the saturation
density $m^\ast_r$ is about $10\%$ down in our calculations against a
decrease of $30-40\%$ in \cite{SHIOMI}. The differences can be surely
assigned to the absence of tensor couplings in our model.

 A comparison with the DSCM calculations of \cite{AGUIRRE} shows
that the main effect of generalizing the nonlinear $\sigma$ interaction
to the vector meson couplings is noteworthy for the effective $\sigma$
and $\omega$ meson masses. Their density dependence is more abrupt in
the present work, and generally speaking the values obtained here are
lower than in Ref \cite{AGUIRRE}.

\section{Discussion and summary}
\label{SEC7}

In this paper we have proposed an effective relativistic hadronic model
inspired in the DSCM to investigate in-medium hadronic properties, in
terms of the baryon isospin asymmetry. The non-linear $\sigma$-nucleon
interaction is generalized to the isoscalar vector, isovector scalar,
and isovector vector channels. Effective medium dependent couplings
arise at the MFA, and a residual interaction with one and two meson
exchange is obtained beyond the MFA. The equation of state (EoS) for
symmetric nuclear matter is softer than the one corresponding to the
DSCM. The symmetry energy coefficient shows an intermediate behavior
between the QHD model and non-relativistic variational calculations.
The asymmetry dependence of the EoS becomes relevant for densities $n
\geq 3 n_0$, and it is emphasized by the contribution of the $a_0
(980)$ meson exchange. Temperature effects in the range $0<T<100$ MeV
are noticeable in the EoS, but moderate in the effective baryon masses.
As a particular manifestation of asymmetric matter we study hadronic
matter with hyperons, in equilibrium against electroweak decay at T=0.
The $\delta$ coupling is the cause of notable modifications in the
population of  hyperons at high densities. Due to the lower
hyperon-meson couplings, relative to the nucleon-meson ones, the
hyperonic effective masses decrease more moderately as the baryon
density increases.

The effective meson masses have been evaluated at T=0 in the RRPA,
including particle-antiparticle finite contributions. The
regularization procedure left undefined parameters $\lambda_s$ and
$\lambda_d$. We have selected numerical values for them, which differ
by one order of magnitude, and are representative of the possible
dynamical regimes. In asymmetric nuclear matter the scalar $\sigma$ and
$\delta$ mesons exhibit monotonous decreasing masses for high
densities, whereas the vector $\omega$- and $\rho$-meson masses show a
slight increase for $n/n_0>1$. In all cases the effective masses remain
below its vacuum values at extreme densities. The dependence on the
asymmetry $\chi$ is more evident for the vector mesons. In
$\beta$-stable hadronic matter the density variation of the effective
masses of all the mesons considered is damped, becoming approximately
constants at twice the saturation density. This effect is due to the
particle-antiparticle terms, contributing even for particles out of
their Fermi shell.

\acknowledgements{ This work was partially supported by the CONICET,
Argentina.}

\end{document}